\documentclass{article}
\usepackage{spconf,amsmath,graphicx}
\usepackage[dvipsnames]{xcolor}
\definecolor{linkcolor}{RGB}{200, 50, 150}
\definecolor{refcolor}{RGB}{100, 50, 100}
\usepackage[
colorlinks=true,
linkcolor=Maroon,
citecolor=BlueViolet,
pageanchor=true,
plainpages=false,
pdfpagelabels,
bookmarks,
bookmarksnumbered,
]{hyperref}
\usepackage{amssymb,amsfonts}
\usepackage{bm}
\usepackage{nicefrac}
\usepackage{subfigure}
\usepackage{caption}
\usepackage{enumitem}
\usepackage{multirow}
\usepackage{tabularx}
\usepackage{orcidlink}

\title{Frequency Disentangled Features in Neural Image Compression}
\name{%
\begin{tabular}{@{}c@{}}
Ali Zafari\,$^\text{\orcidlink{0000-0003-3438-6792}}$\quad 
Atefeh Khoshkhahtinat\,$^\text{\orcidlink{0009-0006-4809-3949}}$\quad
Piyush Mehta\,$^\text{\orcidlink{0000-0002-5240-2322}}$\quad
Mohammad Saeed Ebrahimi Saadabadi\,$^\text{\orcidlink{0000-0003-4112-626X}}$\\ 
Mohammad Akyash\,$^\text{\orcidlink{0000-0001-5187-1269}}$\quad 
Nasser M. Nasrabadi\,$^\text{\orcidlink{0000-0001-8730-627X}}$
\end{tabular}}
\address{West Virginia University, USA}

\begin{document}
%
\maketitle
\begin{abstract}
The design of a neural image compression network is governed by how well the entropy model matches the true distribution of the latent code. Apart from the model capacity, this ability is indirectly under the effect of how close the relaxed quantization is to the actual hard quantization. Optimizing the parameters of a rate-distortion variational autoencoder (R-D VAE) is ruled by this approximated quantization scheme. 
In this paper, we propose a feature-level frequency disentanglement to help the relaxed scalar quantization achieve lower bit rates by guiding the high entropy latent features to include most of the low-frequency texture of the image. In addition, to strengthen the de-correlating power of the transformer-based analysis/synthesis transform, an augmented self-attention score calculation based on the Hadamard product is utilized during both encoding and decoding. 
Channel-wise autoregressive entropy modeling takes advantage of the proposed frequency separation as it inherently directs high-informational low-frequency channels to the first chunks and conditions the future chunks on it.
The proposed network not only outperforms hand-engineered codecs, but also neural network-based codecs built on computation-heavy spatially autoregressive entropy models.
\end{abstract}
\begin{keywords}
Neural image compression, frequency disentangled features, transformer-based transform coding, augmented self-attention
\end{keywords}
\section{Introduction}\label{sec:intro}

Neural image compression (NIC) has been developed into a mature field of study in recent years \cite{ yang2022introduction}. Simultaneously estimating and minimizing the entropy of the latent code is the main focus to make those neural networks capable of outperforming hand-crafted codecs. To do so, compute-intensive context entropy models have been proposed at the expense of lengthy decoding time \cite{minnen2018autoregressive, qian2021globalref, qian2022entroformer}. 

\begin{figure}[t]
    \centering
    \includegraphics[width=0.39\textwidth]{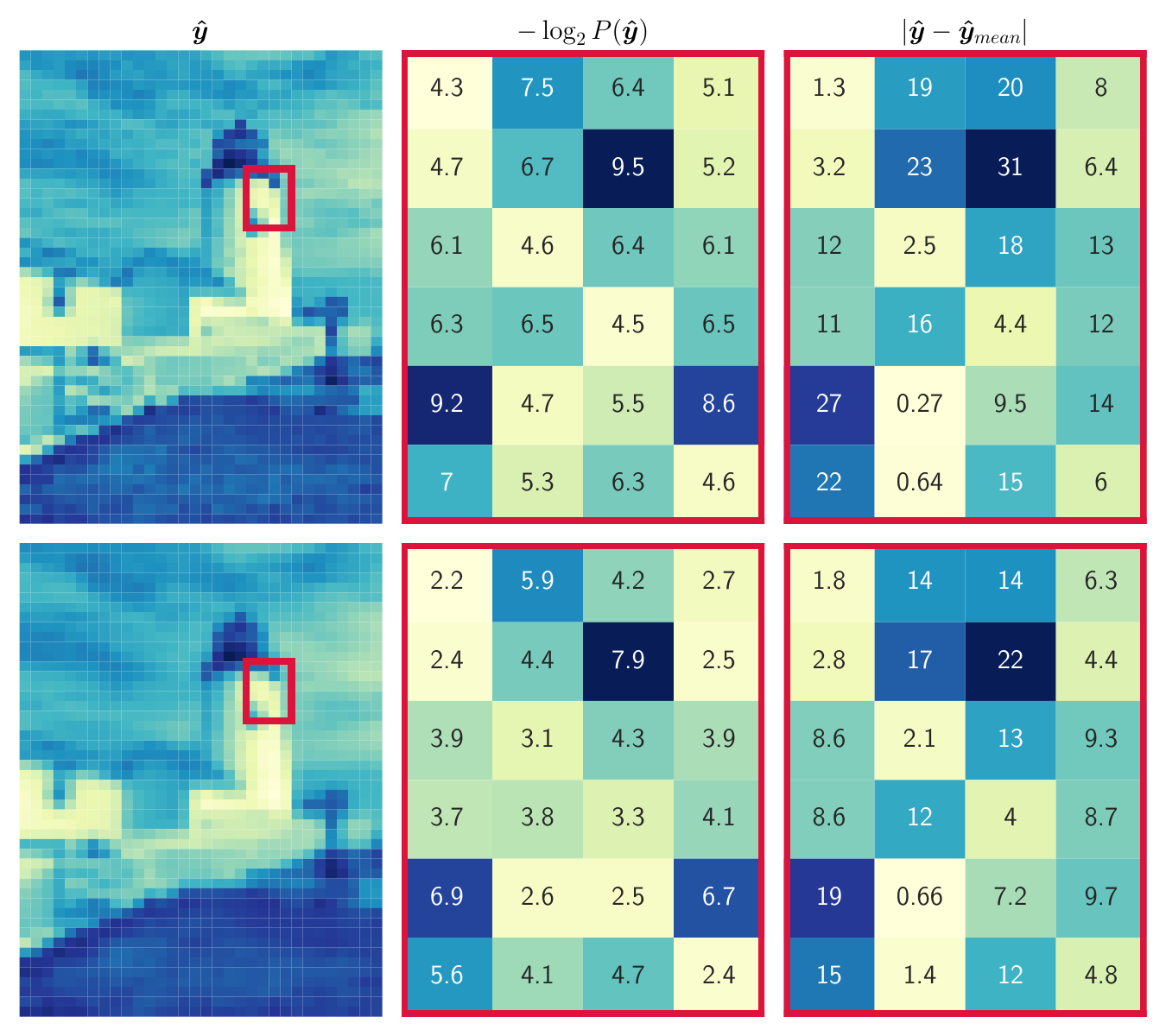}
    \caption{Top row: Transformer-based NIC. Bottom row: Transformer-based NIC with frequency disentanglement. Average of the first chunk of latent code, including the highest entropy channels of sample image from the Kodak dataset. The left column visualizes the whole channel, the middle represents the amount of information of the selected patch and the last column shows the difference from the predicted mean by the entropy model.}
    \label{fig:channel}
\end{figure}

Most of the recent studies rely on a relaxed scalar quantization scheme proposed in \cite{balle2016quantize}. In this approach unit uniform noise is added to the latent representation during training, and gets rounded to the nearest integer during evaluation. We propose to enhance this relaxed quantization scheme by guiding the feature maps of the latent code into high and low-frequency features. Inspired by the quantization scheme in JPEG \cite{jpeg} codec, in which high-frequency non-perceptible details are quantized more coarsely, the relaxed scalar quantization can be improved by disentangling the latent code feature map into high and low frequencies. By doing so, we show that without the need for computation-heavy spatial context entropy models, our proposed model is able to outperform state-of-the-art NIC and hand-engineered codecs.

\begin{figure*}[t]
    \centering
    \includegraphics[width=\linewidth]{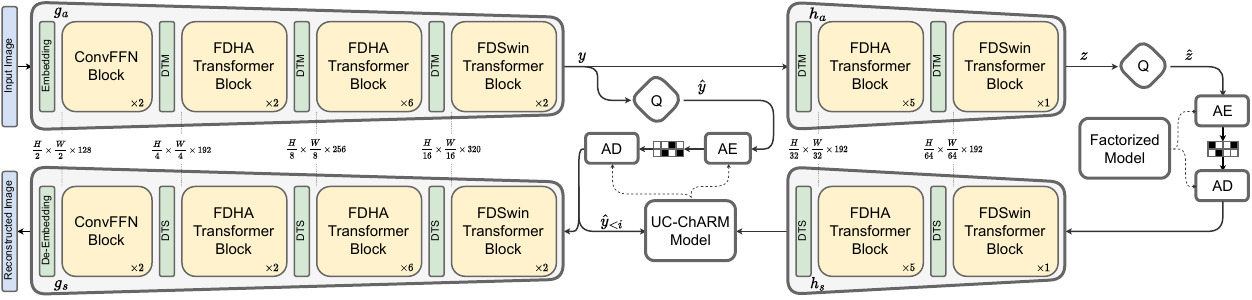}
    \caption{Our network architecture. FDHA and FDSwin are Frequency Disentangled with/without the Hadamard self-Attention module, respectively. DTM/DTS are deformable token merging/splitting \cite{pan2022LITv1}. ConvFFN denotes a layer of depthwise convolution. Entropy models for the latent code are unevenly-chunked channel-wise autoregressive (UC-ChARM). AE/DE and Q refer to arithmetic entropy en/decoding and quantization, respectively.}
    \label{fig:network-arch}
\end{figure*}

In addition, we have addressed the performance degradation of vision transformers as a building block used in NIC \cite{zou2022window, zhu2022transformcoding}. The local self-attention mechanism has been criticized  \cite{zhu2022transformcoding, han2022connection, dai2021coatnet, farahani2023, zhou2021elsa} for hindering the inherent long-range dependency modeling ability of transformers. Authors in \cite{zhu2022transformcoding} also showed that RD-VAE loss will exacerbate this problem, resulting in inferior performance. 
By augmenting the attention score calculation in the local self-attention method, we increase the modeling capacity to address this issue.

Application of Swin \cite{liu2021swin} transformers in NIC was first studied in \cite{zou2022window}, in which the self-attention mechanism appeared only in the autoencoder and keeping the entropy model based on convolutional layers. Fully transformer-based NIC was studied thoroughly by authors in \cite{zhu2022transformcoding}, showing the superiority of transformers in entropy modeling as well. Strengthening the entropy model by spatial context was first proposed in \cite{minnen2018autoregressive}, which let the NIC methods to outperform BPG \cite{bpg} codec for the first time. Built on that, authors in \cite{qian2021globalref} proposed to look for the best spatial context, not necessarily the local context. Their idea was subsequently improved  \cite{qian2022entroformer} by using the top-k most similar contexts using the self-attention mechanism. In another venue of works, to compensate for the slow decoding time of the spatial context models, authors in \cite{minnen2020channelwise} proposed a channel-wise context modeling as a replacement for the spatial context. Grouping both spatial and channel-wise context entropy modeling together \cite{he2022elic}, built a NIC model which outperforms state-of-the-art VVC \cite{vvt2022} codec.
Our contributions are summarized below:
\begin{itemize}[noitemsep,topsep=0pt,parsep=0pt,partopsep=0pt]
    \item By having frequency disentangled heads in the multi-head self-attention, the quantized latents are guided to be separated into local/high frequency and global/low frequency features, which decreases the bitrate by having more precise quantization, as shown in Figure \ref{fig:channel}. 
    \item Criticized local self-attention mechanism of transformer is addressed by augmenting the attention score calculations by Hadamard product-based terms.
    \item Evaluation of our network on high-resolution test images shows that the channel-wise autoregressive entropy model fortified by frequency disentanglement lets our network outperform the state-of-the-art neural and hand-engineered lossy image codecs.
\end{itemize}

\section{Method}
\subsection{Rate-Distortion Variational Autoencoder (R-D VAE)}
Lossy compression is concerned with the tradeoff between rate $R$ and distortion $D$,
\begin{equation}
    R+\lambda D
    \label{eq:rate-distortion}.
\end{equation}
Distortion can be measured by any desired metric, such as mean squared error. Cross entropy between the true distribution and modeled distribution of the latent code represents the rate term in the loss function.

To realize a NIC model with RD-VAE, a learned prior over latent code $\bm{\hat{y}}$, must be shared between the encoder and decoder. This prior can be approximated with a latent variable model called hyperprior entropy model \cite{balle2018hyperprior}. In this model, the latent code is dependent on some side-information/hyper-prior $\bm{\hat{z}}$. The distribution of $\bm{\hat{z}}$ is approximated by a factorized model.
According to Figure \ref{fig:network-arch}, the network input ($\bm{x}$) and output ($\bm{x'}$) relations can be summarized as follows:
\begin{equation}
 \begin{aligned}
    &\bm{x'}=g_s(\bm{\hat{y}};\bm{\theta_g}),\\
    &\bm{\hat{y}}=\lfloor g_a(\bm{x};\bm{\phi_g})\rceil,\\
    &\bm{\hat{z}}=\lfloor h_a(\bm{y};\bm{\phi_h})\rceil,
\end{aligned}
\end{equation}
in which, $\lfloor\cdot\rceil$ denotes quantization. Encoder and decoder nonlinear transforms are represented by $g_a$ and $g_s$ with their learned parameters, $\bm{\phi_g}$ and $\bm{\theta_g}$, respectively. The subscripts $a$ and $s$ refer to \emph{analysis} and \emph{synthesis}.

\begin{figure}[t]
    \centering
    \subfigure[FDSwin transformer block. Local window (W)/shifted-window (SW) self-attention is applied in this module.]{\includegraphics[width=0.42\textwidth]{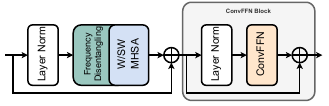} \label{fig:fdswin}}
    \vfill
    \subfigure[FDHA transformer block.]{\includegraphics[width=0.42\textwidth]{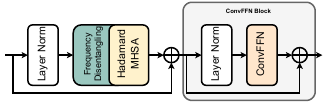}\label{fig:fdha}}
    \vfill
    \subfigure[Frequency disentangle module (left) /  Hadamard augmented self-attention module (right).]{\includegraphics[width=0.49\textwidth]{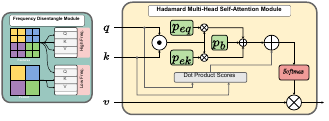} \label{fig:fd_hadamard}}
    \caption{Network architecture building blocks.}
\end{figure}

The loss for the Rate term is a cross-entropy between the true and approximated distribution of latent code derived as:
\begin{equation}
    R=\mathbb{E}_{\bm{x}\sim p_{\bm{X}}}[-\log_2P_{\bm{\hat{y}}|\bm{\hat{z}}}(\bm{\hat{y}}|\bm{\hat{z}};\bm{\theta_h})-\log_2P_{\bm{\hat{z}}}(\bm{\hat{z}};\bm{\gamma})],
\end{equation}
where $\bm{\theta_h}$ and $\bm{\gamma}$ are parameters of the learned entropy model on latent code ($\bm{\hat{y}}$) and hyper-prior ($\bm{\hat{z}}$), respectively. The block diagram of our proposed network is shown in Figure \ref{fig:network-arch} which consists of FDSwin, FDHA, and ConvFFN modules (see Figures \ref{fig:fdswin}, \ref{fig:fdha}). Building blocks of the proposed RD-VAE are discussed in subsequent sections. 

\subsection{Relaxed Quantization}
To make optimizing the parameters of the RD-VAE feasible, the hard quantization is simulated by adding unit uniform noise as proposed in \cite{balle2016quantize}. So the latent code will have a probability density as:
\begin{equation}
    p_{\bm{\Tilde{y}}}(\bm{\Tilde{y}}|\bm{\Tilde{z}})=\prod_i(\mathcal{N}(\mu_i,\sigma_i^2)*\mathcal{U}(-0.5,0.5))(\Tilde{y}_i),
\end{equation}
where $\mu_i,\sigma_i^2$ are the variational parameters and $\bm{\Tilde{y}}, \bm{\Tilde{z}}$ are relaxed versions of $\bm{\hat{y}}, \bm{\hat{z}}$, respectively. These are estimated as a function of hyper-prior $\bm{\hat{z}}$ and previously decoded samples (in case of autoregressive modeling) by the entropy model discussed in section \ref{subsubsec:entropy-model}.

\subsection{Autoregressive Entropy Model} \label{subsubsec:entropy-model}
Entropy modeling jointly on hyper-prior and the context information \cite{minnen2018autoregressive} is the de-facto standard in NIC. The more the entropy model gets closer to the true distribution of the latent code, the less information is required to be transmitted to the decoder. Minnen \emph{et. al.} \cite{minnen2020channelwise} proposed to replace the masked convolution context \cite{minnen2018autoregressive} with a mask over the channel dimension, resulting in faster decoding time and enhanced rate-distortion performance. As there was no priority between the equally divided groups of channels, an unevenly chunked slice of channels is proposed in \cite{he2022elic} to let the most of low entropy channels be dependent on the high entropy ones. We have borrowed the same Unevenly Chunked Channel-wise Autoregressive (UC-ChARM) model in which the channels clustered into $\{16,32,32,64,176\}$ chunks. In this setup, each cluster is dependent on all its previous clusters.
By doing so most of the information related to the global low-frequency details of the image will be captured in the first high entropy chunk of channels (Figure \ref{fig:channel}). Therefore, the variational parameters of the entropy model can be expressed as
\begin{equation}
    (\mu_i,\sigma_i^2) = f_{gp}(h_s(\bm{\hat{z}}),f_{\scriptscriptstyle UC-ChARM}(\bm{\hat{y}}_{<i});\bm{\psi}_{gp}),
\end{equation}
where $f_{gp}$ is a neural network with parameters $\bm{\psi}_{gp}$, outputting the Gaussian distribution parameters conditioned on both the hyper-prior ($\bm{\hat{z}}$) and previously decoded latent codes ($\bm{\hat{y}}_{<i}$).

\subsection{Frequency Disentangled Features}\label{subsubsec:frequency-disentangle}
As the human visual system is less sensitive to minute high-frequency details of the image to be compressed, we will enforce a frequency decoupling in the latent features of the network \cite{chen2019octaveconv, pan2022hilo}. We split heads in the multi-head self-attention (MHSA) module in two. The high-frequency heads look for local details within each window. Then by average pooling the tokens in each window, low-frequency heads capture the global structure on each window as shown in Figure \ref{fig:fd_hadamard}. In the output, both high/low-frequency tokens are concatenated and fed to the next block.
This frequency disentanglement along with the unevenly distributed channels in the entropy model helps the RD-VAE to achieve better performance than the state-of-the-art compute-intensive NIC networks which are based on modeling the serially decoded spatial context \cite{ minnen2018autoregressive,qian2021globalref,qian2022entroformer,he2022elic}.
This frequency separation in self-attention can be written as:
\begin{equation}
    SA(X) = [\underbrace{\text{MHSA}(Q,K_L,V_L)}_{\text{Low Frequency}}\frown\underbrace{\text{MHSA}(Q,K_H,V_H)}_{\text{High Frequency}}],
\end{equation}
where $\frown$ denotes concatenation.
By having the frequency separation in the latent code, the channel-wise autoregressive model will place most of the low-frequency features in the first chunk of channels. This is due to the fact that all subsequent chunks modeled are dependent on this first one \cite{minnen2020channelwise, zhu2022transformcoding, he2022elic}. As a result, the relaxed quantization by adding unit uniform noise is guided to distinguish between chunks of channels. Keeping the most informational features minutely modeled in the first one and letting the high-frequency quantized values appear on the last clusters, occupying fewer bits in the final bit-stream.

\subsection{Augmented Hadamard Self-Attention} \label{subsubsec:hadamard-attention}
Convolutional content-agnostic networks in transform coding, do not respect differences in distributions of the input image. Although efforts have been made to generate filters online during the processing of the image \cite{li2021involution, zhou2021decoupled}, vision transformers exploit a different approach with the help of self-attention.
Window-based local self-attention transformers \cite{liu2021swin} have been studied thoroughly to find a solution for their degraded performance compared with the global computationally expensive version.
Authors in \cite{dai2021coatnet} observed that the performance of local self-attention is almost the same with convolutions, proposing a mixed conv-transformer architecture.
\cite{han2022connection} investigated the similarity between local self-attention and dynamic depth-wise convolution and showed that it performs on par with the local self-attention but with lower computational complexity.
Here we fortify the dot product-based attention score calculation by Hadamard product scores \cite{han2022connection,zhou2021elsa}. This Hadamard-based attention score calculation, as shown in Figure \ref{fig:fd_hadamard}, is a fast and plug-in modification to the conventional self-attention based on inner-product, written as:
\begin{equation}
    v_i' = \sum_{j\in \Phi} \frac{e^{q_i^Tk_j +(q_i\odot k_i)p_{ek}^{o}+p_{eq}^{o}(q_j\odot k_j)+p_{b}^{o}}}{\sum_{j\in \Phi}e^{q_i^Tk_j +(q_i\odot k_i)p_{ek}^{o}+p_{eq}^{o}(q_j\odot k_j)+p_{b}^{o}}}v_j,
\end{equation}
where $o=\text{offset}(i,j)$ specifies the neighborhood around position $i$. $p_{ek},p_{ek}$ and $p_{b}$ are the relative positional embeddings and bias, respectively.
\section{Experiments}
\begin{figure}[t]
    \centering
    \includegraphics[width=0.45\textwidth]{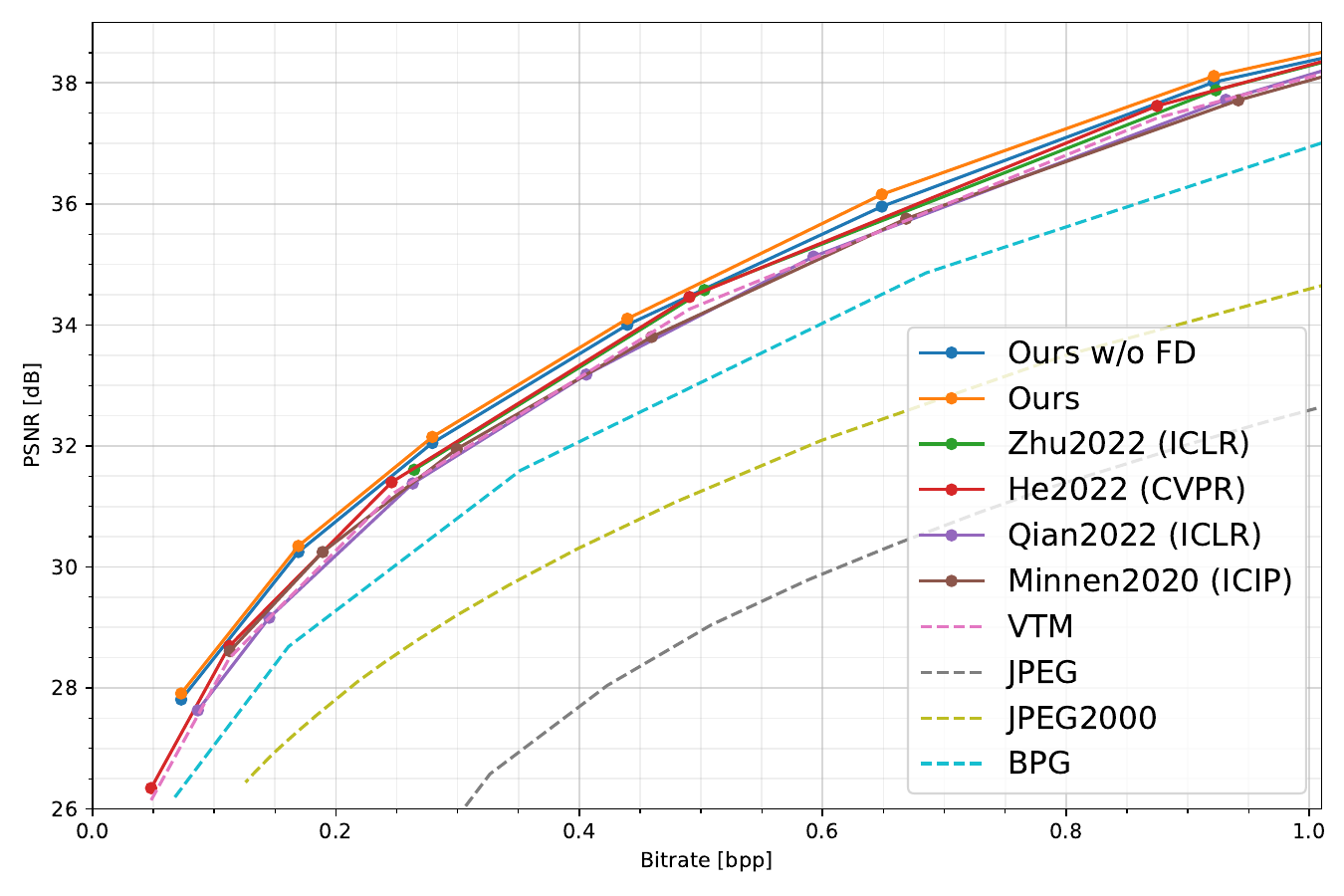}
    \caption{Rate-distortion performance averaged over 24 images of Kodak \cite{kodak} test set.}
    \label{fig:rd-cvrves}
\end{figure}
\subsection{Implementation Details}
To train our NIC network, we have used the combined set of high-resolution images from datasets DIV2K, Flickr2K and CLIC \cite{clic2022}, in a total of 5,285 images randomly cropping $256\times256$ out of them during training. Parameters of all models were optimized using Adam by 2.4M steps with a batch size of 16. The learning rate initial value was set to $1\times10^{-4}$ and gets annealed until the end of training to $1\times10^{-5}$. Hyper-parameter $\lambda$ is chosen from $\{0.0025, 0.0075, 0.015, 0.025, 0.040, 0.055\}$ for 6 different bit-rate regimes. Apart from the rate loss measured by entropy model, mean squared error is used as the measure of distortion.

\subsection{Comparison with the SOTA Methods}
In addition to comparison with hand-engineered codecs, including JPEG, JPEG2000, HEVC-Intra (BPG) \cite{bpg} and VVC-Intra \cite{vvt2022}, we have evaluated our model with respect to state-of-the-art NIC methods. NIC networks are categorized into two groups, spatial context \cite{qian2022entroformer} and channel-wise \cite{minnen2020channelwise, zhu2022transformcoding} entropy modeling. The authors in \cite{he2022elic} model context across both channel and spatial dimensions with a two-pass serial operation. As shown in Figure \ref{fig:rd-cvrves}, our proposed model outperforms all aforementioned methods in terms of rate-distortion.
\subsection{Ablation Study \& Decoding Latency Analysis}
To better investigate the contribution of frequency decoupling in our network, we also trained the same series of networks on different bit rates but without the frequency separation in self-attention modules. Results can be found in Figure \ref{fig:rd-cvrves}, asserting the effectiveness of the proposed frequency disentanglement module.
Apart from the rate-distortion, another important factor for a codec is its decoding latency by which an efficient low complexity decodder is preferred \cite{zhu2022transformcoding, he2022elic, estiri2022low, wiggers2022efficient}. Spatial-dependent context models suffer from slow decoding due to the inherent serial decoding process \cite{minnen2018autoregressive, shoushtari2022prior, mukherjee2022challenges}. A semi-parallel spatial context model is utilized by \cite{ qian2022entroformer,he2022elic} which trades off a reasonable coding performance for faster decoding. As shown in Table \ref{tab:latency} our network enjoys low-latency decoding by having the autoregressive modeling only over the channels instead of spatial locations.
\begin{table}[t]
    \centering
    \caption{Decoding latency compared to state-of-the-art, averaged over Kodak \cite{kodak} test set at an average bit-rate of $0.7$ bpp. (AR: Autoregressive, [p]: 2-step parallel)}
    \begin{tabularx}{0.49\textwidth}{cccc}
    
        \multirow{2}{*}{Model} & \multirow{2}{*}{AR Domain} & \multicolumn{2}{c}{Decoding (ms)}\\\cline{3-4} && $\bm{\hat{z}}$ & $\bm{\hat{y}}$\\
        Minnen \emph{et. al.} \cite{minnen2018autoregressive} & Spatial &  2.1 & $>$5000\\
        \hline
        Minnen \emph{et. al.} \cite{minnen2020channelwise} & Channel &  2.7 & 71\\
        \hline
        Qian \emph{et. al.} \cite{qian2022entroformer} &  Spatial [p] &  5.4 & 96\\
        \hline
        He \emph{et. al.} \cite{he2022elic} & Spatial [p]+Channel &  2.5 & 61.5\\
        \hline
        Zhu \emph{et. al.} \cite{zhu2022transformcoding} & Channel & 3.3 & 82\\
        \hline
        ours & Channel & \textbf{3.4} & \textbf{73}\\
    \end{tabularx}
    \label{tab:latency}
\end{table}
\section{Conclusion}
We proposed a high/low-frequency disentanglement in the latent code for learned image compression. Furthermore, we augmented the self-attention mechanism in transformer-based compression networks. By doing so, our network outperformed recent compute-intensive neural codecs. This superiority comes without deteriorating the decoding latency by relying only on the channel-dependent entropy modeling.\\\\
\textbf{Acknowledgment.} This research is based upon work supported by the National Aeronautics and Space Administration (NASA), via award number 80NSSC21M0322 under the title of \emph{Adaptive and Scalable Data Compression for Deep Space Data Transfer Applications using Deep Learning}.
\bibliographystyle{IEEEbib}
\small{
\bibliography{refs}
}

\end{document}